%% file: ms.tex
\documentclass{article}
\usepackage[utf8]{inputenc}
\usepackage[colorlinks=true, allcolors=blue]{hyperref}
\usepackage{graphicx}
\usepackage{caption}
\usepackage{url}
\usepackage{amsmath}
\usepackage[gen]{eurosym}
\usepackage{subfig}
\usepackage{lineno}
\usepackage{soul}
\usepackage{natbib}
\providecommand{\keywords}[1]
{
  \small	
  \textbf{\textit{Keywords---}} #1
}

\title{Unsub Extender: a Python-based web application for visualizing Unsub data}
\author{Eric Schares\\Engineering \& Collection Analysis Librarian\\
Iowa State University\\eschares@iastate.edu\\ORCiD 0000-0002-6292-8221}
\date{}

\begin{document}

\maketitle


\input{1Abstract}

\vspace{1em}

\input{2Intro_rev1}

\input{3Significance_rev1}

\input{4Funct_rev1}

\input{6demo_rev1}

\input{7Conclusion_rev1}

\bibliography{ms.bib}

\end{document}

%% file: 1Abstract.tex
\section{Abstract}

This article introduces \textbf{Unsub Extender}, a free tool to help libraries analyze their Unsub data export files.

Unsub is a collection development dashboard that gathers and forecasts journal-level usage metrics to provide academic libraries with deeper measurements than traditional cost-per-use.
Unsub gives libraries richer and more nuanced data to analyze their subscriptions, but it does not include a way to easily visualize the complex and interrelated data points it provides.

\textbf{Unsub Extender} (\url{https://unsubextender.lib.iastate.edu}) is a free Python-based web application that takes an Unsub export file and automates the creation of interactive plots and visualizations. The tool loads with example data to explore, and users upload their specific Unsub file to quickly populate the pre-made plots with actual data. 

Graphs are interactive, live-updating, and support zoom, click-and-drag, and hover. Filters are specified through sliders to model scenarios and focus on areas of interest. A drop-down menu allows users to change a journal's decision status, and  graphs update automatically. After evaluating journals, users can export the modified dataset to save their decisions.

Unsub Extender proposes best practice in analyzing the increasingly common Unsub export file. It simplifies the analysis, eliminates duplication of effort, and enables libraries worldwide to make better, more data-driven decisions.

%% file: 2Intro_rev1.tex
\section{Introduction} 

A core mission of the academic library is to provide patrons with access to scholarly materials. This is often accomplished through journal subscriptions with academic publishers which cost millions of dollars per year. 
Scientific journals are critical to support the research and teaching missions of the university and provide an avenue for researchers to communicate their own findings, advancing scientific knowledge and their career progression. 

The ever-increasing price of journal subscriptions, coupled with library acquisition budgets that rise modestly year over year (if at all), has led to a ``serials crisis" in the profession. This community-termed realization dates back to at least the late-eighties \citep{SerialsCrisis1989,SerialsCrisisDouglas,SerialsCrisisRoth, serialscrisis}. 
Journal subscription prices increased over 550\% from 1986 to 2018 (most recent data available), far outpacing inflation which rose 130\% during that same period. Total library expenditures rose 220\% over that time, though this category includes not only serials budgets but also salaries, operating expenses, monograph purchasing, and other expenses \citep{ARL_inflation2, ARL_inflation}. These runaway journal price increases occurred even as journals were moving to an electronic format where the marginal production costs drop to near zero \citep{nature_truecost}. 

The average single-year increase over this 32-year time period was 6.1\% per year for journal prices, 3.5\% per year for total library expenditures, and 2.65\% per year for inflation. Focusing only on the past 10 years (2009-2018), library budgets have remained flat or decreased seven times, while journal prices have continued to climb at 2.5x the rate of inflation during that time (3.36\% journal prices vs. 1.38\% inflation vs. 0.76\% library total expenditures). This constant pressure on library budgets has caused libraries to look very critically at their journal subscription practices.

\subsection{Big Deal} 
The term ``Big Deal" was coined in the mid-1990's by Kenneth Frazier \citep{DilemmaBigDeal} and refers to the practice of publishers bundling hundreds of electronic journal subscriptions together into one large package for a single price. Libraries pay a lower fee than they would by subscribing to each journal in a publisher's list individually, but they are also locked into agreements to pay for titles they may not need or want \citep{PoynderBigDeal}. 
Introduced as a solution to the problem of ever-increasing journal prices, initial Big Deal bundle pricing was generally based on a library's current print subscription spend with the publisher, plus an extra 5-15\% to enable access to non-subscribed journals in the publisher's portfolio and a 6\% inflationary increase per year \citep{pub.1013220935}. 


Several studies have thoroughly investigated the value (or lack thereof) libraries get from Big Deals. 
\citet{Shu_IsItSuchBigDeal} studied journal subscription data from 34 University Libraries and used references to each journal from scholars affiliated with the universities as an indicator of their actual use. The paper found the value of the Big Deal investment to be low, as the university community cited only a fraction of the journals purchased by their institution through the bundle. 
More journals does not necessarily mean more value if the extra journals are not being used, and the ratio of cited journals to all subscribed journals decreased by 61\% between 2000 to 2011. 

Big Deal pricing is also not transparent, consistent, or easily comparable among universities, as each research institution negotiates with the publisher separately and contracts are often shielded under non-disclosure agreements. 
Freedom of Information Act (FOIA) requests and state-level open records laws have been used to reveal Big Deal pricing despite this confidentiality \citep{Knox_openrecordslaws}. Studies show this pricing opacity results in little consistency, with prices varying among institutions and commercial publisher prices between 3-10 times higher than corresponding prices from nonprofit publishers \citep{pub.1013220935, Brundy} 

Commercial academic publishing is a tremendously profitable business. RELX, the parent company of the world's largest academic publisher, Elsevier, reported profit margins of in excess of 30\% from 2016-2019, with 2020 dipping slightly to 29.2\% \citep{RELX}. Much has been written about the unusual working relationship between researchers, universities, and publishers - scientists conduct research, write articles, and give it to publishers for free, provide peer-review to validate other scientists' papers, only for publishers to turn around and then sell the articles back to the university for millions of dollars \citep{nature_truecost, ProfitableBad}. 
One analysis suggests there is enough money this system for a large scale shift away from subscriptions altogether and toward a universal Open Access model \citep{DisruptingMPDL}. 

The top publishers continue to increase their share of scholarly output. \citet{Oligopoly} report that over 50\% of all journal articles published in 2013 and indexed by the Web of Science database were published by only five companies: Elsevier, Wiley, Springer Nature, Taylor \& Francis, and SAGE. 
This is consistent with a wider analysis of WoS article content from 2014-2018, which shows these five companies responsible for 55\% of article output \citep{DataDescriptor}. 
This consolidation is a result of the creation of new journals each year and the ever expanding acquisitions of existing journals by these publishers \citep{Oligopoly, Exclusion}. 
\citet{Shu_IsItSuchBigDeal} used economic theory to explain publishers' motivation to launch new journals each year, noting ``academic publishers increase sales volume and profits by selling additional journals with no use or little use at the maximum price libraries are willing to pay." 
Indeed, \citet{Exclusion} looked at journal publishers as monopolies with an eye toward antitrust law, concluding ``substantive antitrust issues" were present with Big Deal bundling.

\subsection{Unbundling the Big Deal} 
As a result of yearly price increases and the realization of dubious value, some libraries have moved to cancel their Big Deal and opted instead for an a-la-carte model, selecting and paying for each individual journal separately. This practice of breaking up, or unbundling, the Big Deal is becoming more common; 
SPARC, the Scholarly Publishing and Academic Resources Coalition, maintains a database of libraries worldwide who have broken up their Big Deal \citep{SPARCBigDealTracker}. 
The self-reported tracking tool reports 80 such events back to 2004, with 33 occurring between 2019 and August 2021 (institutions may be counted in more than one event if they report multiple Big Deal breakups). 
While unbundling ensures a library pays only for content they have selected, it does impose costs both in terms of higher prices per journal (as the library is now paying the ``list price" and not spreading the cost of the journal across an overall subscription package) and in terms of staff time and effort to curate, assess, and decide on the list of chosen journals on a yearly basis. 

The literature contains many case studies of universities who have unbundled. 
\citet{UnbundlingPractice} take a theoretical approach to studying the evaluation and decisions made when unbundling a Big Deal package, and 
\citet{EJournals_and_BigDealReview} provided a practical review of several papers which deal with cancellation and renewal decisions. The study looked at criteria used in breaking up Big Deals, decision strategies, and impacts of cancellation projects. It found dwindling support for Big Deal packages, with cost-per-use, download statistics, overlap analysis, subject specialist input, and citation analysis found to be present in many of the papers included in the review. 
\citet{BigDealCancelR1} describe the process, data, and communication strategies of breaking up the Big Deal at three R1 Universities, while \citet{ISUBigDealArcand} detail the breaking up of the Springer Big Deal at Iowa State University. 
\citet{LeavingElsevierBigDeal} describes the Italian National Institute of Health's experience in unbundling with Elsevier. They found that cutting 90\% of titles in the package led to only a 22\% reduction in downloads, with half the remaining downloads occurring in 25\% of journals and the top 10 titles (5\%) responsible for 20\% of the downloads. 

In practice, most collection analysis takes place in a homegrown tracking sheet or database, with large, unwieldy files that can be difficult keep up to date short of massive data refresh effort each year. These local spreadsheets require manual research and loading of relevant data points such as citations or authorships, but in practice this extended data may only be added for targeted titles in a package and not all titles. 
Serial decisions at Virginia Tech were made with the help of the ``Big Ugly Database" or BUD \citep{crisis}, a spreadsheet with 90,000 rows of journal titles to make cancellation decisions. \citet{EJournals_and_BigDealReview}'s review also found metrics were commonly fed into a rubric or decision grid that had to be maintained and cleaned, something that takes an extensive amount of staff time. 
More detailed usage data is often acknowledged to be useful, but the difficulty in looking up each journal title individually and adding a single number to the main tracking spreadsheet makes it very cumbersome to do in practice.

\subsection{Unsub.org} 
\label{sec:unsub} 
In order to assist libraries in their evaluation and possible cancellation of Big Deal packages, it needs to be easier to assemble data and see the value of individual titles and the package as a whole. 
Unsub.org \citep{unsub_site} is a collection analysis dashboard that helps academic libraries more deeply analyze their journal subscription packages and Big Deals. Launched in November 2019, the tool has seen rapid uptake in the academic library community and is now used by over 500 libraries on six continents, including national consortia in the UK, Canada, and Australia \citep{piwowar_grant}.

Unsub goes beyond traditional cost-per-use metrics and brings in other measures of demand, such as citations (how often have researchers from an institution cited each journal) and authorship (how often have researchers from an institution published in each journal). 
Users upload COUNTER-compliant electronic usage information, and Unsub uses Microsoft Academic Graph, and soon OurResearch's replacement tool OpenAlex \citep{OpenAlex}, to search, collect, and project the citation and authorship data for the institution over each of the next 5 years \citep{unsub_help_citesauth}. 

A Weighted Usage measure then combines downloads, citations, and authorships from the specific institution to a given journal title $(A)$ into one summary number. 
The weights can be customized within Unsub, with the default weights set as: 

\begin{equation} 
WeightedUsage_A = downloads_A + 10*citations_A + 100*authorships_A 
\label{eq:weighteduse} 
\end{equation} 

The approach of creating a new metric by combining some components which a subscription pays for, like downloads, with some which is doesn't, like authorships, does lead to some important considerations to be aware of. 
Journals have many paths to achieve a high Weighted Usage depending on the specific combination of downloads, citations, and authorships. The fact that two journals may have a similar Weighted Usage does not mean they necessarily perform identically; this is why Unsub Extender encourages a deeper analysis and provides views that go further than relying only on this top-line, summary usage number. 
Choosing between a journal with high publication activity but few downloads or citations or a low publication journal with many downloads and citations can be difficult and will likely necessitate bringing in other data points which Unsub provides (see Section~\ref{sec:IFpercentage}). 

The relative weights reflect the fact that publishing an article in an academic research journal is a more time- and resource-intensive effort than simply downloading an electronic copy of a paper from that journal. 
It is true that publishing a paper in a journal does not require a researcher's library to provide current read access, but the selection of a publication venue reflects careful consideration by researchers and as such, authorships are weighted as equal to 100 downloads. Citations are also a valuable signal of usage, and are weighted between downloads and authorships, worth 10 downloads a piece. 

This combined technique and the default relative weightings assigned by Unsub are comparable to an approach taken by 1Science \citep{1Science}, a similar collection analysis tool that was bought by Elsevier and discontinued \citep{1Science_dead_Elsevier}. 
1Science used a dynamic weighting scheme that reflected the comparative statistics for a given journal compared to all titles in the package, with the weighting for authorships equal to the ratio of all downloads to all authorships and the weighting for citations equal to the ratio of all downloads to all citations. 
This approach led to higher absolute weighting numbers, and a larger relative weighting ratio. 
In a limited analysis where the two approaches were compared directly on a sample data set, 1Science weightings came out to 16 and 550 for citations and authorships, respectively. 
These factors result in higher overall ``usage" numbers, but the relative performance of each journal within the package was still roughly equivalent to that of Unsub. 

In addition to demand (as Weighted Usage), Unsub also models alternative modes of instant fulfillment, such as Open Access availability (papers published free to read without a paywall). 
The share of papers published Open Access continues to increase, with estimates that over half of articles published since 2015 are now available through some form of OA \citep{Piwowar_stateofOA}, and 70\% of all article views are projected to be to OA articles by 2025 \citep{Piwowar_futureofOA}. 
In addition, Unsub considers a library's backfile or post-termination retention rights; that is, older content a library has already paid for and would retain even without a current year's subscription. 
Together, these modes of fulfillment show how much of the institution's usage can be met even without a current year's subscription to the journal, going beyond traditional cost-per-use analysis and giving a more nuanced assessment of the cost-effectiveness of the title. 

Unsub then projects usage five years into the future to give a sense of what to expect for each journal title. The usage, citation, and authorship numbers are modeled as demand, and Interlibrary Loan, Open Access, and backfile are modeled as modes of fulfillment. 
Bringing these disparate data points together into one unified dashboard illuminates the actual value of a journal's subscription and helps libraries make decisions on what to keep and what to cancel. 

Since its launch in 2019, Unsub has garnered positive attention and use cases from the library collection development community, and it continues to make news as libraries use this data to break up Big Deals and unsubscribe to large journal packages \citep{UnsubReviewCharleston, UnsubScience, UnsubSPARC}. 
While this approach may be useful for analyzing journal packages offered by many publishers, Unsub currently supports analysis of five: Elsevier, Wiley, Springer Nature, Taylor \& Francis, and SAGE, the same five publishers that accounted for half of 2013's article content in Web of Science \citep{Oligopoly} and 55\% from 2014-2018 \citep{DataDescriptor}. 

%% file: 3Significance_rev1.tex
\section{Significance of this Project} 
So far, we have seen that libraries are motivated to break up their Big Deals, and in many cases have no choice due to continually rising prices and stagnant budgets, but they are confronted with difficult data collection and analysis workflows. 
Unsub was then developed to provide a way for libraries to more easily analyze their subscription packages and investigate which journals are providing value for their patrons, bringing multiple modes of demand and projections of fulfillment together into one place. 

\begin{figure}[h] 
\centering 
\includegraphics[width=1\textwidth]{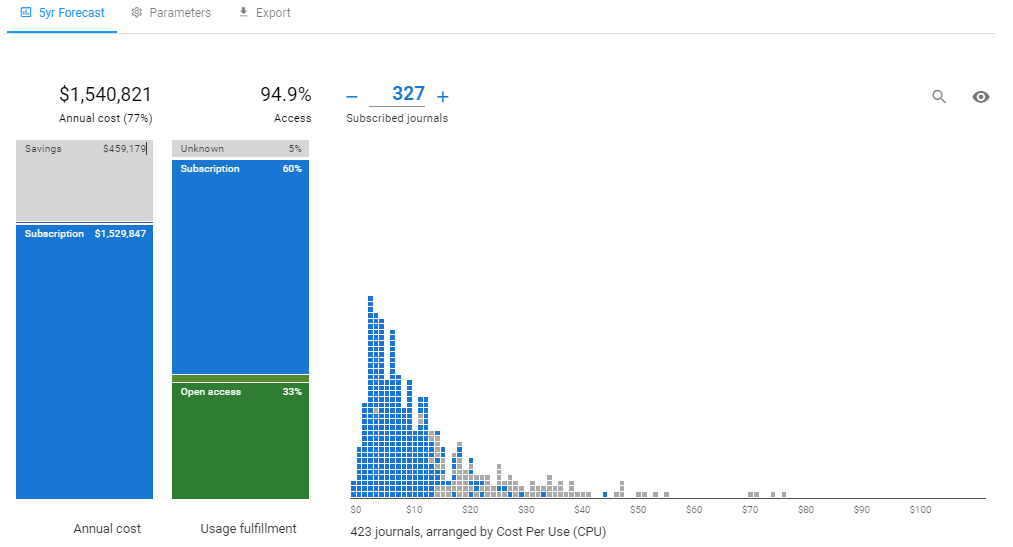} 
\caption{Summary graph provided by Unsub} 
\label{fig:unsub_graph} 
\end{figure}

Unsub's strength is that it supplies libraries with very rich data on the individual journal level to help them analyze the value of subscriptions, but its weakness is that it does not help users easily understand the many complex and interrelated data points it provides. In fact, Unsub only includes one visualization: a high-level summary at the overall package level (Figure \ref{fig:unsub_graph}). In this graph, each small square box is an individual journal title the user can select and toggle to subscription status on or off; blue squares are marked as titles to keep, gray squares are those that will not be renewed. 
This graph is useful to glean an overall summary of a publisher's Big Deal package as a whole, but it does not help the user make a decision on any individual title in particular. 

Clicking on a journal title's square brings up a display box with more data about the title, but it is difficult to compare numbers among similar journals or decide where a cutoff should be applied. Furthermore, it is not clear to see the underlying data which is driving the location of the square along the x-axis - the title may have many downloads but few authorships, or may be cited frequently but not published in often (see Equation \ref{eq:weighteduse}). 
Unsub does provides an alternative to the graph it calls ``Table View" which shows some information on multiple journals at once, but it is limited in size, requires lateral scrolling, and cannot display all the collected data on each journal title. 

The graph in Figure \ref{fig:unsub_graph} shows Iowa State University's actual decisions on a large commercial publisher's subscription package that was analyzed with Unsub. It is clear that all retained (blue) journals do not form a continuous group; in practice, some are located far along the x-axis toward high ``Adjusted Cost-per-Use" but were still kept due to local reasons (traditional strengths, support of a small department, strategic university emphasis area, etc). 
The scattering of decisions along this metric demonstrates a need for deeper analysis and systematic thinking about which journals to keep than Unsub has provided. 

Therefore, a new tool called Unsub Extender was developed and is formally introduced here.


\subsection{Unsub Extender} 

\textbf{Unsub Extender} is a Python-based web application which takes a standard Unsub \texttt{.csv} export file as input and automates the creation of interactive plots and data visualizations \citep{UnsubExtender}. 
These graphs help libraries better understand their Unsub data so they can model different scenarios, make more informed cancellation decisions, and view those decisions in the wider context of the overall journal package to help assess if their conclusions are consistent. 
It is freely available at \url{https://unsubextender.lib.iastate.edu}, and the source code is openly posted at the project's GitHub repository \citep{UE_Github}. 

\begin{figure}[h!] 
\centering 
\includegraphics[width=1\textwidth]{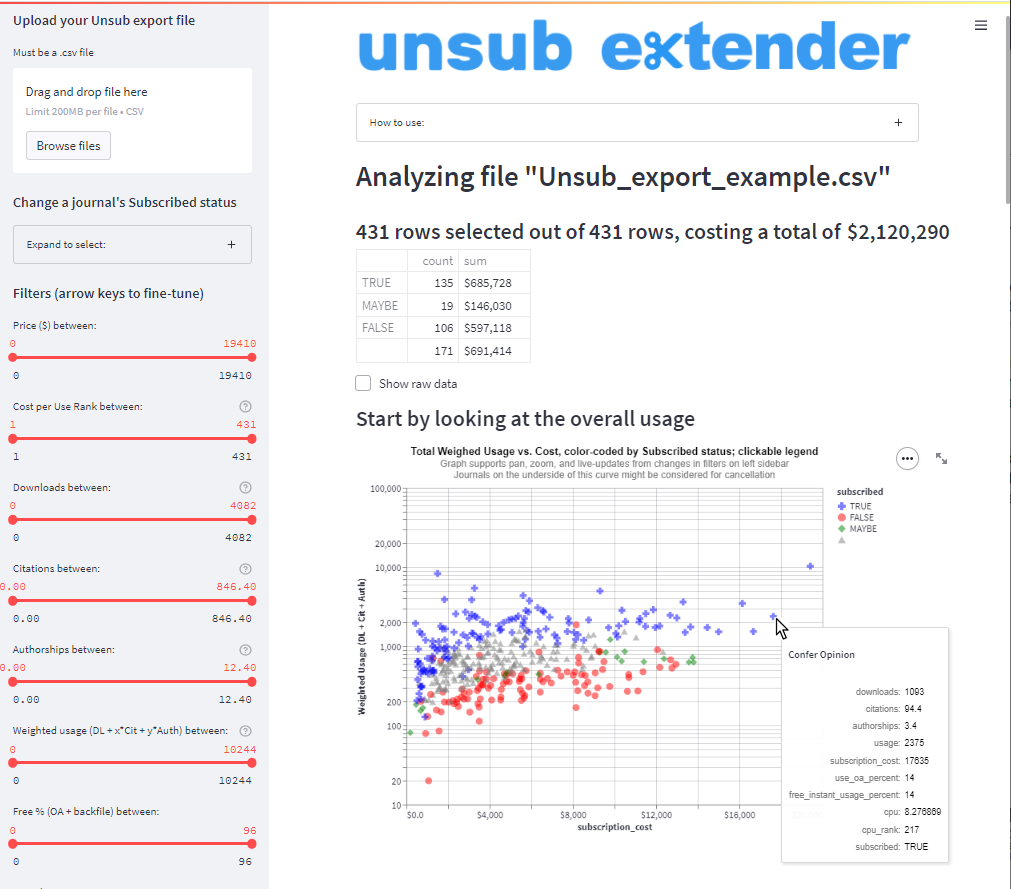} 
\caption{Unsub Extender main screen} 
\label{fig:full_screenshot} 
\end{figure}

The tool launches with a sample Unsub dataset so users can explore the set of twelve pre-made standard plots, or they may upload an actual data file to quickly populate the graphs with their live data. 
Figure \ref{fig:full_screenshot} shows a screenshot of the main landing page of the site, with the logo, summary data table, left-hand filtering sidebar, and scatter plot of Weighted Usage (from Equation \ref{eq:weighteduse}) vs. subscription cost for each journal. 

All graphs are interactive, responsive, live-updating, and support zoom, pan, and click-and-drag. 
Hovering over a data point shows more information about the journal title; for example, Figure \ref{fig:full_screenshot} shows the hover box with Unsub's detailed information for the fictional journal, ``Confer Opinion," including number of downloads, citations, authorships, usage, cost, and more. 
The graph makes it clear to see the overall trends and groupings of titles to keep or cancel that emerge (more detailed discussion of the available graphs will occur in the next section). 

A summary data table above the graph shows the running total of decisions so far. It includes both the absolute number of titles and the cumulative dollar amount in each Subscription category of TRUE, MAYBE, FALSE, or -blank-, making it easy to see where the analysis stands at any given time, how different scenarios would play out, or how close a package is to a desired dollar amount goal. 
Filters for various criteria are specified through sliders on the left panel to help model different cut-off levels and focus on areas of interest, and 
a drop-down menu allows users to change a journal's ``Subscribed" decision status and the graphs will then update accordingly.

\subsection{Contribution} 

Unsub Extender aims to solve the problem of understanding the rich data Unsub provides, enabling better collection development decisions to be made more quickly. 
The motivation for creating this tool came from the author's job responsibilities, which include running the Unsub tool, understanding the report, summarizing it, and communicating the conclusions to the collection management team. 
It took a substantial amount of work to internalize the various aspects of the detailed Unsub data, experimenting with columns to plot against each other, calculating custom extended measures, and deciding how to color-code and present the graphs. 
Eventually, through trial-and-error the collections team arrived at a satisfactory set of graphs that helped decide which journals to retain and which to cancel, but the question remained: if we were having this much trouble understanding and digesting the Unsub data, what about the other 500 libraries around the world that \emph{also} subscribe to this tool \citep{piwowar_grant}? 

Unsub data files are becoming more widespread as academic libraries look to unbundle their journal subscription packages, but a standardized way to analyze the Unsub export file has not yet been proposed. 
Unsub Extender is intended to encode the author's expertise and assist other libraries with their journal analysis procedures. 
Using the pre-made graphs that have already proven useful speeds up others' analysis and enables the wider library community to more quickly gain understanding of the overall journal package value, as well as that of each individual journal. 
By proposing best practice in how to analyze the increasingly common Unsub data file, the tool shares what we have learned locally, enables peer libraries to benefit from our work, eliminates duplication of effort, and allows more informed decisions to be made more quickly.

%% file: 4Funct_rev1.tex
\section{Functionality} 

To demonstrate the available graphs and the functionality of Unsub Extender, sample data is set to pre-load in the application and automatically populate upon visiting the site. 
This sample data set is based upon a real publisher package and its subsequent performance at Iowa State University, but with identifying details such as ISSN stripped out, possible identifying data such as price and usage slightly modified, and journal titles changed to invented, fictional names. Using a genuine data set as the starting point helps demonstrate what actually happens when Unsub Extender is used on an authentic Unsub data export file and models where the trends and bands of decisions begin to emerge. 

This sample data shows the state of a journal package at the \textbf{conclusion} of an analysis project, with cancellation decisions previously made. These demonstrate how decisions generally fell within bands of renewal vs. cancellation when looking at the graphs. 
Some journals had no explicit decision made about them (those in gray triangles) because the target savings goal was reached, and those were renewed by default. 
In an actual journal analysis scenario, data points coming in would be set to FALSE and would initially load as a uniform color, gradually changing as renewal decisions get made on a title-by-title basis. 

The user is able to dynamically make changes to the status of individual journal titles through the Unsub Extender site; however, changing the Subscribed status of more than a handful of titles at one time can become cumbersome in the tool. 
A more likely and efficient approach is to edit the underlying data file itself, filtering and sorting based on different criteria and changing the Subscribed status of multiple titles at once. 
The data file can then be loaded back into Unsub Extender, the graphs plotted and updated, and a visual representation of the renewal decisions created. 

A suggested workflow is an iterative process: making decisions on journals, reviewing the trends and groupings, making further decisions, and adjusting where needed. 
Unsub Extender is intended to help users both identify journals for cancellation and also assess and visualize how those decisions are working out. 
For example, if a group of titles is decided to be kept, similarly performing data points graphed near them might also be marked for renewal. These similarities can be difficult to visualize when looking at the exported data file as a spreadsheet. 
These graphs may reveal inconsistencies in the decisions, help the team assess similar titles without a decision yet, or make trends apparent which may (or may not) have been intended. 
A useful exercise for a visitor to the site may be to continue the decision making process and analyze those journals with a (blank) status. Which would you recommend to keep or cancel? 

A key aspect enabling the creation of this project is the policy of Unsub's developers, OurResearch, to remain responsive to the research community and make their projects open and extensible, which enables users to develop new features or capabilities \citep{OurResearch}. 
One result of this commitment, though by no means the sole benefit, is Unsub's ability to export the data very easily once it has been calculated. 

\begin{figure}[h] 
\centering 
\includegraphics[width=.99\textwidth]{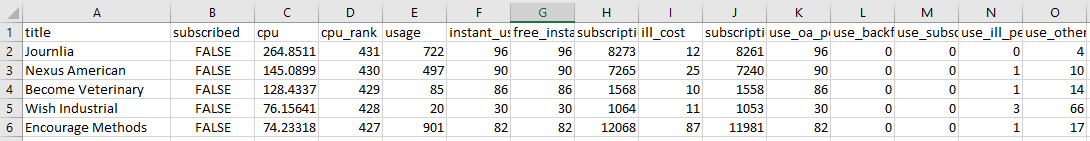} 
\caption{Sample exported Unsub \texttt{.csv} file} 
\label{fig:export} 
\end{figure}

Figure \ref{fig:export} shows an example of the raw exported Unsub \texttt{.csv} file, with each row containing a range of calculated metrics for an individual journal title. While this detailed data is very interesting, it is difficult to grasp the meaning behind all 35 columns of data for each journal title. Unsub provides a summary to explain the columns \citep{unsub_help}, but the overwhelming amount of data presented in this format makes it difficult to use in making decisions. 

Uploading the exported data to Unsub Extender populates the twelve graphs that are coded and running in the site, designed to show different angles of the underlying figures. 
Using the predictable and standardized Unsub export file format means the site can be assured of the incoming data structure, what heading names are used, and what columns contain which data types. 
Arriving at the same conclusion when looking at a data set through multiple graphs provides confidence that the decisions are sound and can be defended. It can also illuminate mistakes or inconsistencies in the treatment of data points, as a journal marked for cancellation that is surrounded by other journals of similar performance being kept will raise the question of why that journal is treated differently from its neighbors. 

A brief overview and explanation of the graphs will now be provided. 

\subsection{Weighted Usage vs. Cost} 
\label{subsec:WU_vs_Cost} 
\begin{figure}[htbp] 
\centering 
\includegraphics[width=.9\textwidth]{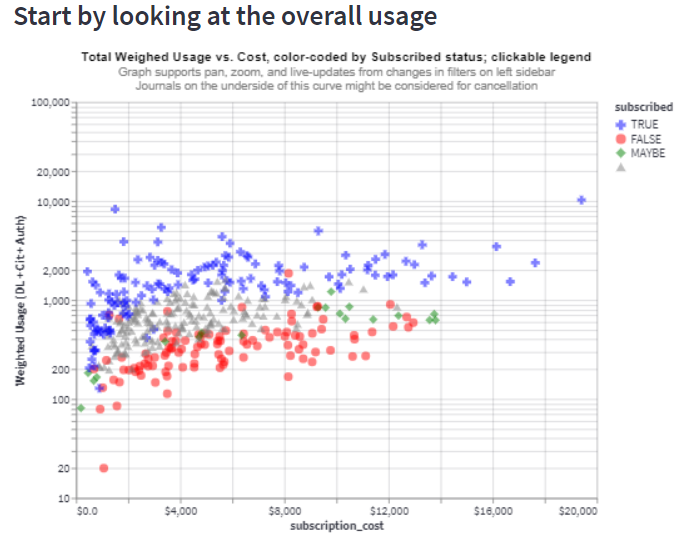} 
\caption{Color-coded by Subscribed status} 
\label{fig:graph1} 
\end{figure}

\begin{figure}[htbp] 
\centering 
\includegraphics[width=.9\textwidth]{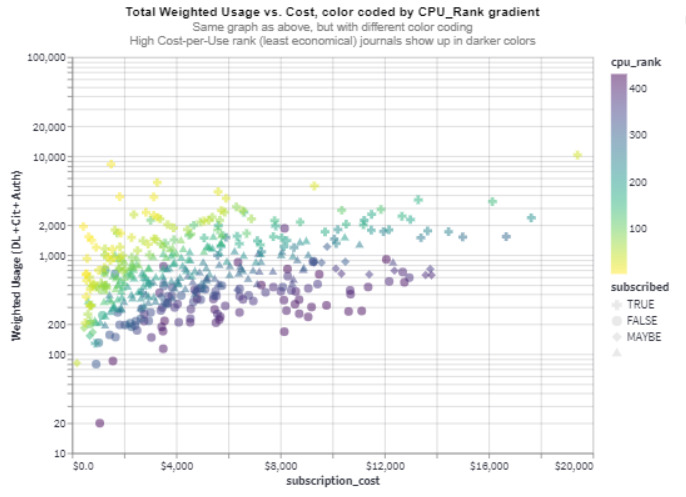} 
\caption{Color-coded by CPU rank} 
\label{fig:graph2} 
\end{figure}

The first two graphs that display when loading Unsub Extender show Weighted Usage vs. Subscription Cost (Figures \ref{fig:graph1} and \ref{fig:graph2}). These visualizations quickly emerged as the central decision making bodies in actual journal cancellation projects, as they provide a suitable balance between detailed information and a high-level summary. The first two graphs actually plot the same data points but use different color codings: one is by Subscribed status (Figure \ref{fig:graph1}), the other reflects each journal's relative cost-per-use (CPU) strength among all titles in the package (Figure \ref{fig:graph2}). 

In addition to color coding the data points by Subscribed status or CPU rank, Figures \ref{fig:graph1} and \ref{fig:graph2} also encode the Subscribed status decision by symbol, with KEEP's showing as a plus sign, FALSE titles as a circle, MAYBE's as a diamond, and titles without a declared decision set to a triangle. This adds a layer of data encoding to provide an additional layer of detail for those who may not easily perceive color changes. 
The Subscribed symbols in the legend are all clickable and gray out any non-selected categories, allowing for more focus on a particular category. 
It is possible to select multiple categories by holding the Shift key and clicking, while using the Subscribed status radio button filter in the left-hand pane will show a single category at a time. 

Looking first at Figure \ref{fig:graph1}, it becomes clear that decision bands and patterns emerged in the previously analyzed data. Journal titles appearing on the underside of the curve were targeted for cancellation (red) since their price does not support their usage. 
These titles have a low value compared to others in the package; alternatives can offer more usage for the same amount of money (looking up vertically), or less money for the same amount of usage (looking left horizontally). 
Titles which appear on the top of the curve have been marked as blue for renewal. These titles \emph{do} provide value for the subscription cost, as the Weighted Usage is among the highest in the package and cancellation would be unlikely. 

Sharp-eyed readers may notice a red circle among the middle of the blue band at \$8,000 and 2,000 Weighted Usage in Figure \ref{fig:graph1}; more will be said about this journal (code-named ``Science Advance") and the reasons for its cancellation decision later in Section \ref{sec:Demo}. 

A few green diamonds are scattered throughout the middle band of Figure \ref{fig:graph1}. These are points that were marginal, on the border between cancellation and retention. They were set as ``MAYBE" in Unsub Extender's terms, intended to be marked for conversation at a later date when coming back to the data. They may be the first to be considered for cancellation should additional cuts be needed to meet a budget goal, or scrutinized further with a different set of stakeholders and perspectives. 
Finally, the remaining journals have had no decision made about them, and these appear as triangles colored in background gray. 

Figure \ref{fig:graph2} shows the same data and is useful to compare journals to others within the package. The color-coding is now by gradient of CPU rank, something Unsub calculates and orders from 1 to ``number of journals in the set" (431 in this case). Again we see that the least economical journals (those with highest CPU rank) tend to appear on the bottom of the curve, mostly displaying as circles (FALSE's) in darker shades of purple and blue. Light yellow represents the lowest CPU journals, or most economical, with the general trend of light to dark following the trend of blue to red in Figure \ref{fig:graph1}.

\subsection{Individual demand metrics} 
With a basic understanding of the overall distribution and shape of the dataset, the next set of graphs take the user deeper into investigating each component of Weighted Usage individually (Equation \ref{eq:weighteduse}). 

\begin{figure}[h] 
\centering 
\includegraphics[width=.99\textwidth]{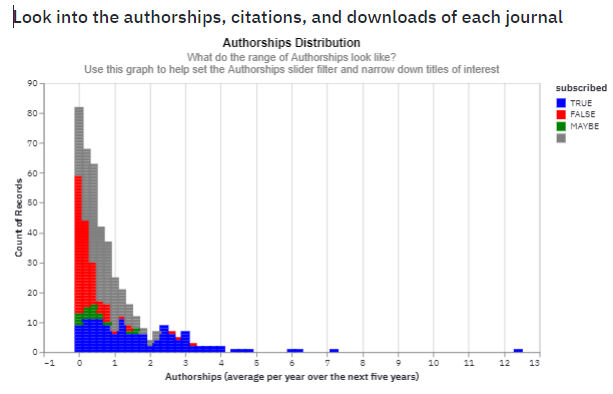} 
\caption{Authorship count distribution} 
\label{fig:graph3} 
\end{figure} 

Figure \ref{fig:graph3} isolates authorship data and displays the histogram distribution for the entire journal package. Authorship data is the most heavily weighted component of Weighted Usage, but it is often very skewed. 
In this dataset, most journals are projected to have three or fewer authorships per year for the next five years, but about 4\% of the titles project to have more than 3, and one high outlier title has over 12. 
A high authorship number is often reason enough to keep a journal, as cancellations of low Weighted Usage titles will by definition have a low number of authorships. 

\begin{figure}[h] 
\centering 
\includegraphics[width=.99\textwidth]{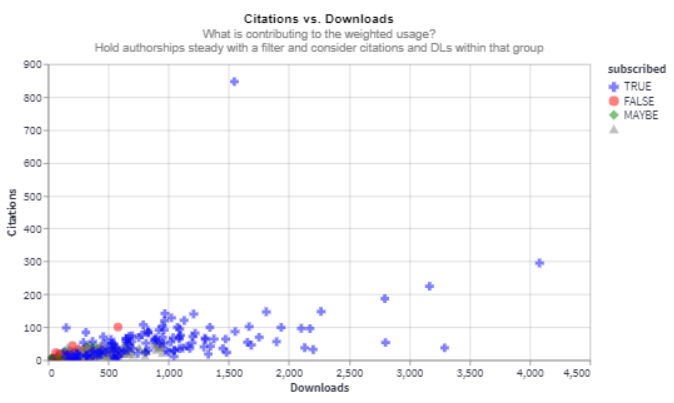} 
\caption{Citations vs. Downloads} 
\label{fig:graph4} 
\end{figure} 

Figure \ref{fig:graph4} focuses on the other two components of Weighted Usage, citations and downloads, and plots one against the other in a scatter plot. There is a weak positive relationship between the two; as downloads tend to rise, so too do citations. 
One clear outlier in this dataset is the journal title with many more citations than would otherwise be expected at over 800 (shown near the top middle of Figure \ref{fig:graph4}). This title, ``Citing Practice," is projected to be heavily cited by the research institution, and as such, may be interesting to investigate further. Low usage journals in the lower-left hand corner are generally marked for cancellation, though they can be difficult to see in this static screenshot. The live tool would allow the user to zoom and pan to investigate further if desired.

\begin{figure}[h!] 
\centering 
\includegraphics[width=.99\textwidth]{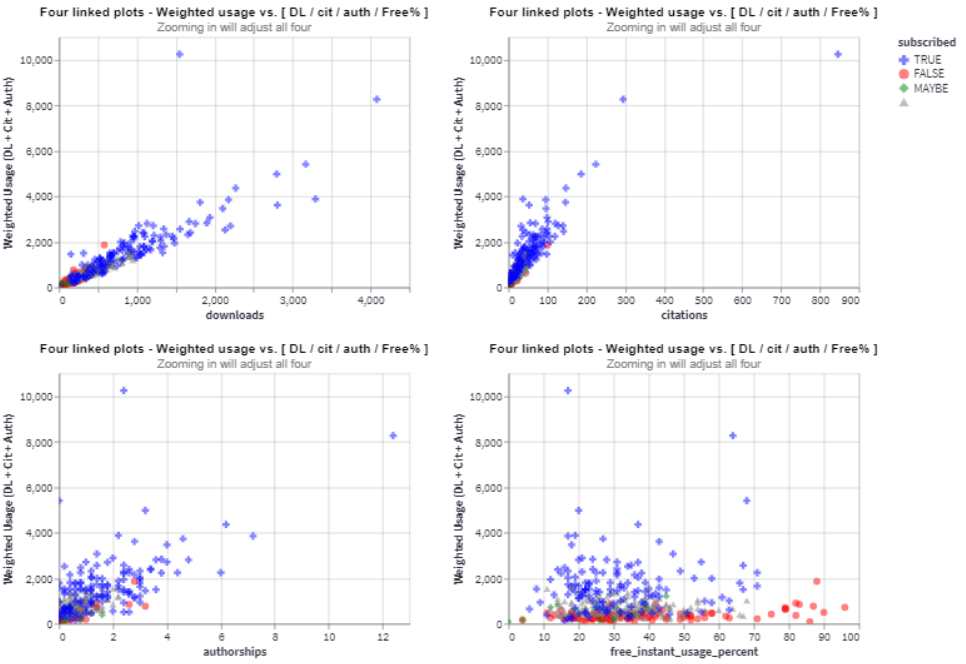} 
\caption{Four linked plots - Weighted Usage vs. downloads, citations, authorships, and Open Access percentage} 
\label{fig:four_linked_plots} 
\end{figure} 

Next, Figure \ref{fig:four_linked_plots} shows Weighted Usage plotted against each of its three components individually and introduces a new measurement. The four plots in Figure \ref{fig:four_linked_plots} are all linked to each other on the site, which means that zooming in or panning around on one plot will execute the same action on the other three connected plots. 
Breaking out the Weighted Usage calculation and plotting separately against each of its components allows for users to more deeply understand what is driving this summary metric. For example, it becomes evident that the outlier data point up around 10,000 Weighted Usage, the highly cited title ``Citing Practice," is not driven by many downloads, or even a large amount of authorships, but by the largest number of citations in this group of journals by far. This then translates to the highest overall Weighted Usage in this set of 431 journals. 

The bottom-right plot in Figure \ref{fig:four_linked_plots} shows Weighted Usage vs. Open Access percentage. This OA\% measurement is one of several fulfillment methods provided by Unsub to estimate how much content in each journal is freely available 
even without a paid subscription. 
Unsub's help files define OA\% as an estimation of ``the percentage of usage that can be fulfilled via Green, Hybrid, or Bronze Open Access" \citep{unsub_help}. Unsub also notes that users can choose to exclude some of these access types in their dashboard projections if they wish; Bronze OA, for example, may be considered an unreliable mode of access for some institutions as it can be removed by the publisher at any time. (More information on the Open Access colors and how Unsub defines them can be found at the Unpaywall website, a service also run by OurResearch \citep{Unpaywall_OAtypes}). 

In this example, most journals with very low Weighted Usage continue to be marked for cancellation, but those with OA\% over 70\% are also marked to cancel even as Usage begins to rise. In that region, Usage may be rising but Open Access availability rises as well. Making decisions on this criteria does require more work to understand exactly how the journal is providing Open Access content, and an institution must make sure they are comfortable with relying on that mode of fulfillment before they mark the title for cancellation.

\subsection{Instant Fill Percentage} 
\label{sec:IFpercentage} 
The Instant Fill percentage (IF\%) is an advanced measure of how much value a journal's current year subscription actually holds. Unsub refers to this important IF\% in passing, but it is not explicitly calculated or provided to the user in the exported \texttt{.csv} file. 
An example of Unsub's use can be seen earlier in Figure \ref{fig:unsub_graph}, where it shows a total 94.9\% ``Access" rate even with the decision to cut 25\% of the journals in the package for a savings of \$459,000. 
The IF\% measures how much a current year's subscription to each individual journal contributes toward the blue ``Subscription" bar as part of the total Access rate.

Some portion of researcher demand will still be met even without a current subscription to a particular title. 
IF\% takes this into account by considering other routes of access to that journal's material, such as Open Access availability or backfile access to older issues the library purchased in previous years. If demand for a journal can be satisfied by Open copies or through older years' material, a current year's subscription is less cost-effective. In other words, IF\% measures how much of a journal's usage can \emph{only} be met by having a current subscription and how much each individual journal contributes to the current year's usage of the overall package as a whole. 

Unsub Extender calculates the IF\% for each journal individually to see which titles rely on an active subscription to fill their demand and contribute to the overall Access rate. 
To calculate Instant Fill percentage for a given Journal A, we first find the percentage of that journal's usage which requires a current year subscription (Equation \ref{eq:currentyear}). 
Once the current year's usage for Journal A is found, it is divided by the $total\_weighted\_usage$, the sum of all journals' Weighted Usage in the package (Equation \ref{eq:IF}): 

\begin{equation} 
\label{eq:currentyear} 
CurrentYear_A = \frac{100 - (OA\%_A + Backfile\%_A)}{100}* Usage_A 
\end{equation} 

\begin{equation} 
\label{eq:IF} 
IF\%_A = \frac{CurrentYear_A}{total\_weighted\_usage} 
\end{equation}

\begin{figure}[h] 
\centering 
\includegraphics[width=.9\textwidth]{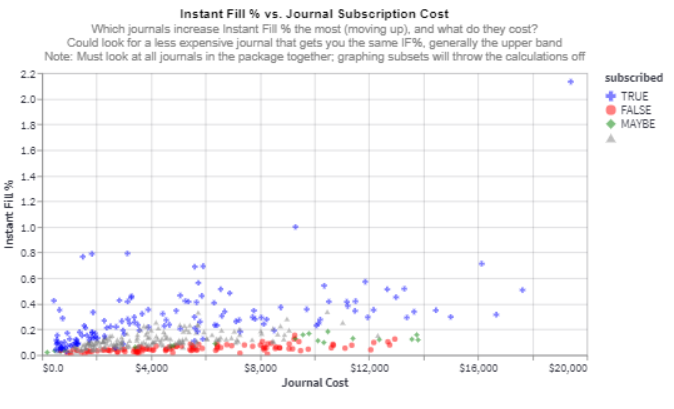} 
\caption{Raw Instant Fill \% vs. Cost} 
\label{fig:IF_single_1} 
\end{figure} 

\begin{figure}[h] 
\centering 
\includegraphics[width=.9\textwidth]{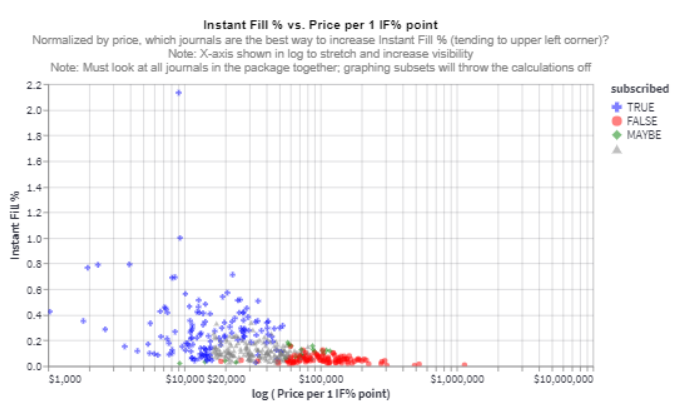} 
\caption{Normalized Instant Fill \% vs. Cost} 
\label{fig:IF_single_2} 
\end{figure}

Figure \ref{fig:IF_single_1} shows the calculated IF\% contributions for each journal title vs. the title's subscription cost. 
Individual journal titles contribute IF\% values in the range of 2\% or less for the example package of 431 titles; that is, one particular journal is responsible for 2\% of an entire package's current year demand (the aforementioned very highly cited journal shown in the top right corner of Figure \ref{fig:IF_single_1}). 
Journals to consider for cancellation are those that appear on the bottom edge of the data, low on the y-axis (low IF\% contributions). If given a choice between two journals that are vertically separated, the title higher on the y-axis would be a better choice to keep, buying more Instant Fill percentage points for the same amount of money. 

To look at another angle of the data, Figure \ref{fig:IF_single_2} normalizes the IF\% for all journal titles to show how much it would cost for that journal to ``buy" a standardized 1\% point of contribution. For example, if the top journal contributes 2.1 IF\% points and costs nearly \$20,000, it would normalize to \$10,000 for one IF\% point. 
This approach is intended to enable comparisons among titles which cost widely different prices and contribute widely different amounts of IF\%. Under-performing journals on this plot appear in the lower-right hand corner, meaning their cost is high (to the right on the x-axis) for a small amount of benefit. (Note: the x-axis on Figure \ref{fig:IF_single_2} is shown in $log$(price) to stretch the axis and increase visibility). 

A final note to consider is that while interacting with IF\% graphs on the website by zooming and panning is safe, running samples and subsets of Unsub .csv files through Unsub Extender will cause problems and miscalculations when it comes to looking at IF\%. The system is only able to see the journal data that is loaded into the tool, and the $total\_weighted\_usage$ number in the denominator of Equation~\ref{eq:IF} will be thrown off if only some portion of the package is included in the analysis, skewing IF\% and making these graphs unreliable.

\subsection{Histogram and Subject Areas} 

\begin{figure}[h!] 
\centering 
\includegraphics[width=.9\textwidth]{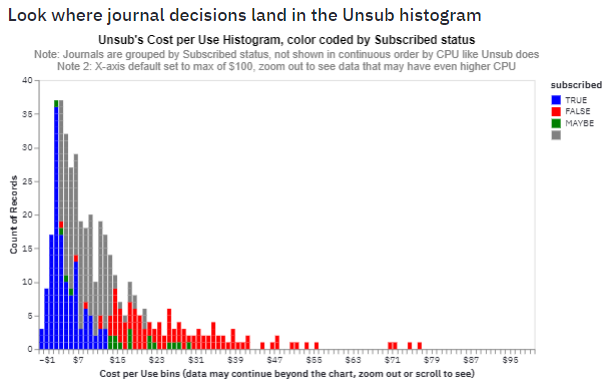} 
\caption{Histogram of titles by CPU, similar to Unsub's provided graph} 
\label{fig:boxes_histogram} 
\end{figure}

The last set of graphs are intended to serve as a final check once cancellation decisions have been iteratively made. First, there is an approximate reproduction of Unsub's summary graph which shows journal titles as boxes color-coded by Subscription decision status (Figure \ref{fig:boxes_histogram}). 
This is intended to provide continuity back to the Unsub tool, but offers more granularity in supported color-coding and fine-tuning for various cancellation scenarios. Red cancelled titles tend to appear in higher Cost per Use bins, though there are some exceptions. 

\begin{figure}[h!] 
\centering 
\includegraphics[width=.99\textwidth]{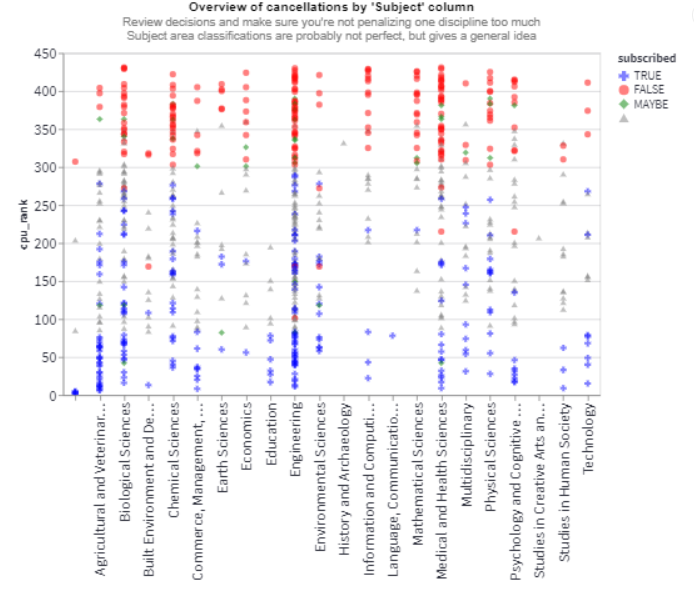} 
\caption{Journals by subject area} 
\label{fig:subjects} 
\end{figure}

Finally, special care needs to be taken to ensure that cancellation decisions are not disproportionately impacting a certain subject area or discipline. Figure \ref{fig:subjects} shows the journal titles in the package graphed by CPU rank and subject area, color-coded by Subscription decision status. 
These broad-level subject area assignments are provided by Unsub and use the codes assigned to each journal by OpenAlex, an open catalog of metadata from the same creators as Unsub \citep{OpenAlex_STI}. 
Prior to summer 2022, the subject codes were assigned by the organization Excellence in Research for Australia (ERA) \citep{AustraliaERA}, and the Unsub Extender code does have conditions to detect and support this older subject data. 

As with any attempt to classify journals into specific disciplines, it is important to recognize that interdisciplinary titles are difficult to classify and should be interpreted with caution, as there is always the possibility of overlap among several subjects that do not get completely captured in the encoding. 
The newer ERA assignments commonly classify a single journal into multiple categories, though it is still non-exhaustive. When this happens, Unsub Extender repeats the assignment and displays the same journal in multiple subject areas. 

As an example, in Figure \ref{fig:subjects}, we can see journals in this package classified as Economics have many of them marked for cancellation and two marked as MAYBE. Viewing the data from this angle may nudge the green points back to safety to avoid impacting a specific discipline too harshly. 
It is also worthwhile to consider whether a given title also appears in another subject area; in this case, one of the green Economics titles also appears in the Commerce \& Management subject area (the green point at 300 cpu\_rank), while the second green Economics title has a single assigned class. 

\subsection{A Note of Caution} 
Looking strictly at usage vs. cost graphs or making decisions on data points because they appear in a certain area of the chart is an attractive but simplified and reductive view of the complex procedures involved in collection analysis. 
Treating data as simple points on a graph can be useful for an initial high level look, but it is key to remember the critical importance of applying subject-matter expertise in cancellation activities and not depending solely on numbers. 
Applying human judgement to consider other data not represented such as department size, understaffed or emerging research areas, or intangibles like strategic initiatives to the university can help avoid mistakes in cancelling journal titles. 
After all the hard number crunching and pattern detecting is done, there is still no substitute for an actual human using their expertise and experience for one final look.

\section{Left Hand Panel} 
The left side of Unsub Extender consists of fly-out panel with control features that include the Upload button, Subscribed status editor, sliders to filter the data, and Export button. 

\subsection{Upload} 
Unsub subscribers can export their own unique \texttt{.csv} data file, then choose the \textbf{Browse files} button in the top left corner to actually upload their real data. All graphs and tables will then automatically overwrite and update to reflect the real Unsub measurements, automating the analysis and clearly showing which journal titles may be considered for cancellation. Uploaded user data does not get saved within the tool, and once a user closes the browser tab the data is lost.

\subsection{Subscribed} 

The \textbf{Subscribed} column within the exported \texttt{.csv} file is where decisions about the renewal status of each journal are recorded. 
Unsub Extender uses that information to color-code the data points in many graphs, and the definitions that it supports are: 

\begin{itemize} 
\item TRUE (blue plus): a journal to continue subscribing to 
\item FALSE (red circle): a journal to cancel 
\item MAYBE (green diamond): a journal to highlight or mark to think about later 
\item ::blank:: (gray triangle), a journal with no decision assigned yet 
\end{itemize} 

\begin{figure}[h!] 
\centering 
\includegraphics[width=.4\textwidth]{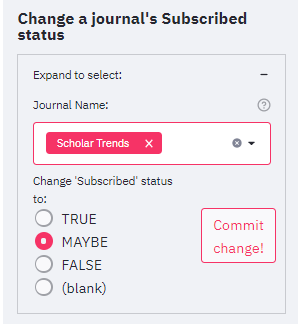} 
\caption{Subscribed status editor} 
\label{fig:subscribed_editor} 
\end{figure}

The first two categories are consistent with Unsub's internal structure and follow those conventions, while the second two choices are added by Unsub Extender itself for further flexibility. 
As users explore the Unsub data, it is possible to dynamically change the Subscribed status of a journal using the editor in the left-hand panel. Opening the Expand attribute reveals a text box to search for a journal name, and the radio buttons below allow a choice of status to move to. Figure \ref{fig:subscribed_editor} shows the journal ``Scholar Trends" being changed to a MAYBE. 

\subsection{Sliders} 
After an overview of the graphs, a user may want to filter the data to more deeply focus on a certain area of interest. 

Dual-sided sliders can be seen in red on the left-hand side of Figure \ref{fig:full_screenshot} and allow for easy investigation and filtering based on measures such as price, cost-per-use rank, downloads, citations, authorships, weighted usage, Open Access percentage, OA + backfile percentage, and Subscribed status. 
Changing any of the criteria will automatically update all graphs and tables as well, with only those data points matching the specified criteria included in subsequent analysis. Filters work as a cumulative AND, so setting multiple filters and criteria reduce the data points to those which satisfy \emph{all} conditions.

\subsection{Export} 
After users explore, modify, and decide on journal subscription states, the Export option will create a new copy of the data file with decisions saved and updated. This can then be used for further analysis, to re-upload to Unsub Extender in the future, or to save for decision records. The button will download a time-stamped file named in \texttt{YYYY\_MM\_DD-HH\_MM\_SS.csv} format.

%% file: 6demo_rev1.tex
\section{Demonstration of Unsub Extender} 
\label{sec:Demo} 
To demonstrate how Unsub Extender facilitates the exploration of Unsub data, let us follow a single journal title through the process of evaluation with the provided graphs. 
The fictional title ``Science Advance" was briefly referred to earlier in the paper (Section \ref{subsec:WU_vs_Cost}) as a title that had been marked for cancellation, even though its calculated measurements placed it among other titles that were being renewed. 

Figure \ref{fig:graph1} shows this red circle in a sea of blue, located at \$8,000 subscription cost and 2,000 Weighted Usage. Hovering over the data point on the actual Unsub Extender site would show modest downloads and fairly strong citations and authorships, consistent with a healthy Weighted Usage number. However, in Figure \ref{fig:graph2} we see the CPU rank assigned by Unsub to be 405 out of 431, meaning it is in the bottom 6\% of all journals in the package in terms of total cost effectiveness. 

(One tip to follow a single title through the set of graphs is to use the Subscribed status editor (Figure \ref{fig:subscribed_editor}) to temporarily mark the title as MAYBE; this turns the data point to a green triangle in all the visualizations and makes it easier to track from chart to chart.) 

The linked plot showing Weighted Usage vs. OA\% (lower right of Figure \ref{fig:four_linked_plots}) reveals the reason for its low current year value. This journal has an estimated Open Access availability of 88\%, meaning nearly 9 in 10 projected ``uses" could be filled for free even without a current year's subscription. 
The Instant Fill Percentage graphs also demonstrate this title is located among others that have been marked for cancellation, as it contributes only 0.06\% to IF and has a very high Normalized IF\% at \$134,000 per 1 point of IF\%. 

This example shows why solely using the traditional cost-per-use, or even the more sophisticated Weighted Usage, is not always enough to make informed decisions. Bringing in alternate modes of fulfillment provides additional context and clarity and completes the picture to help a library understand if it really is cost effective to continue subscribing to this journal.

\subsection{Python Libraries and Hosting} 
Finally, a few details on the technical aspects of Unsub Extender are provided. 
Unsub Extender is written in Python, which offers thousands of open source libraries to extend its built-in functionality. 
The tool uses the Python libraries Altair and Streamlit for its structure. 

Data visualization is an active area of Python development with hundreds of plotting libraries to choose from \citep{PyViz}, and it was a non-trivial process to narrow down the many libraries to arrive at a graphing library that would support the vision of this project. 
Altair is a ``declarative statistical visualization library for Python, based on Vega and Vega-Lite" \citep{Altair} and provides the underlying code for all the graphs. 
Altair supports interactivity (zoom, pan, hover) and builds on top of the simple Vega-Lite declarative calls, with a clear grammar and many worked examples to build from. The interactivity was a crucial component, as other common libraries such as Matplotlib render as static images and therefore would not support the interactivity that was required to enable hover, pan, and zoom. 


Streamlit is a library that ``turns data scripts into shareable web apps" \citep{Streamlit}. Streamlit provides the structure to turn Unsub Extender's Python script into an actual web application and enables the addition of filters, sliders, and file uploading. Streamlit offers free basic web hosting for initial projects, but the capacity was quickly overrun with demand when even a few people accessed the site simultaneously. Therefore, while the site continues to use Streamlit's web components and features, hosting is now provided by Iowa State University's College of Liberal Arts and Sciences Information Technology Department.

%% file: 7Conclusion_rev1.tex
\section{Conclusion} 

This paper has formally introduced the tool \textbf{Unsub Extender} and demonstrated how it can assist libraries in analyzing journal subscription data provided by the collection analysis tool, Unsub. 
Exploring the rich Unsub data can be intimidating, but Unsub Extender allows an academic institution to simply upload their exported data to fill in the pre-made graphs for a plug-and-play experience. 
The tool quickly provides insight into the many complex and interrelated data points from Unsub, and it guides users through the analysis to quickly arrive at consistent, understandable, and defensible conclusions. 

Unsub Extender proposes best practice in analyzing the increasingly common Unsub data file, and it is intended to assist libraries around the world in making informed collection development decisions and extracting maximum value from their read-access journal subscriptions. 
In short, Unsub provides data that can help libraries break up their Big Deal, and Unsub Extender helps translate that data into graphs to make the analysis more easily digestible to see how breaking up the Big Deal may affect the institution.

\subsection{Reception} 
Launched in May 2021, Unsub Extender has seen over 2,000 hits from 37 countries.
Direct feedback from users has been limited but positive. 
The majority of overall traffic has come from the United States, but recent months have seen a shift in traffic to a majority from outside the United States, with a large amount of usage from the UK. 
The site was named a 2022 winner of the RUSA ETS Best Emerging Technology Application (BETA) award by the ALA Reference and User Services Association, Emerging Technologies Section \citep{RUSA_ETS_BETA_Award} and the 2022 winner of the Innovation in Access to Engineering Information Award by the Engineering Libraries Division of the American Society for Engineering Education \citep{ASEE_ELD_Award}.

Unsub and the team at OurResearch have been enthusiastic supporters of Unsub Extender, promoting the tool in a Tweet \citep{Unsub_UE_tweet} and providing a link to the site directly from the Unsub Export tab itself. 
The author delivered a joint presentation with Unsub co-founder Heather Piwowar on case studies of how Unsub is used in real life \citep{NASIG}, and a detailed walk-through and live demonstration of the tool was given as part of a series of Unsub webinars \citep{UE_webinar}.

\subsection{Limitations and Future Work} 
Unsub Extender's release is stable and new or upgraded features form the basis of future work. 
Enabling analysis of multiple supported publisher packages at the same time is a possible area of future development. This currently works by uploading one combined Unsub file; however, the resulting analysis is complicated as there is no way to segment by publisher that would also support a single-publisher use. 
Ranked calculations would also be affected as the cpu\_rank lists for each publisher package will introduce duplicates of each number. 
In addition, translating the usage instructions to offer versions in multiple languages would also be a helpful development. 

Feedback on the project is always welcome. If users develop a new type of graph at their institution, more graphs could be coded and added to the site for future use. 
Learning which graphs are proving useful to actual users and how other libraries use their Unsub data would be helpful. 
Testing has been done on data sets from the author's institution, but at this time it is unknown how well Unsub Extender would scale when packages become very large (1,000s of journals at once). Soliciting more users, sample data, and feedback would help test the effectiveness of the tool and may reveal better or more straightforward ways of doing things. 

This project depends on several external tools. First are the Python libraries pandas, Streamlit, and Altair, and careful testing is done after any new versions or releases of these packages to be sure they remain compatible with the project before upgrading on the live site. A \texttt{requirements.txt} file in the GitHub repository specifies certain versions of the libraries and holds them steady to prevent new releases from deploying before they can be thoroughly tested. 

Unsub Extender is also heavily dependent on Unsub. 
At its launch, Unsub used data from Microsoft Academic Graph (MAG) to analyze usage patterns and provide projections over the next five years. With the news that MAG has stopped updating its data source as of December 31, 2021, the team at OurResearch has developed a new, free and open data source called OpenAlex which functions as a drop-in replacement for MAG \citep{MAGout}. 
OpenAlex launched in beta at the beginning of 2022 \citep{OpenAlexAbout}, and Unsub will soon begin using data from this source when calculating metrics. The change in data source has no impact on Unsub Extender, since this project relies completely on Unsub to produce the data that Unsub Extender then interprets and graphs. 
Additionally, Unsub may add new columns to the data export file in the future, but this would also not impact Unsub Extender as long as the required columns and expected names are still present. 
Any changes to these required columns would need to be tracked and reflected in the logic of the site, and the Unsub product development owner is aware of this dependency and is in frequent communication with the author. 

The tool is hosted by the Iowa State University College of Liberal Arts and Sciences IT Department, which provides support for the College's research groups and departments. Any change to the hosting infrastructure which negatively impacted Unsub Extender would also impact many other research projects. 
It is also possible that the author may become unwilling or unable to maintain this project. Were that to happen, the site may indeed fall behind Unsub's export standards and no longer be able to plot the data. 
However, one major advantage of openly posting the project's source code is that in such a scenario, it is conceivable that an interested member of the community could choose to use the code to re-establish and continue with this work. 

In conclusion, Iowa State University has used the information from Unsub and Unsub Extender to help redirect funds away from under-performing subscription journals in large commercial packages 
and re-purpose it toward supporting many developing Open publishing models which align with the library and university's stated principles \citep{ISU_OA_agreements, ISU_OA_principles}. 
The author hopes that future work will simply involve Unsub Extender helping more libraries free up money in their acquisition budgets and accelerate the transition toward Open Access.

\subsection{Acknowledgements} 
The author wishes to thank the team at OurResearch for being supportive of the idea and development of Unsub Extender. 
Thanks also to Nick Booher and the Iowa State University College of Liberal Arts and Sciences Information Technology Department for providing robust web hosting and support. 

\subsection{Conflicts of Interest} 
The author declares no conflict of interest. 

\subsection{Funding Information} 
The author declares no funding. 

\subsection{Data Availability} 
The source code for Unsub Extender and the sample dataset used to demonstrate the functionality are freely available on the project's GitHub repository, licensed under a GNU Affero General Public License v3.0 \citep{UE_Github}.